\documentclass[10pt]{article}
\usepackage[utf8]{inputenc}
\usepackage{graphicx}
\usepackage[most]{tcolorbox}
\usepackage{hyperref}
\usepackage{amsmath}
\usepackage{amssymb}
\usepackage{listings}
\usepackage{color}
\usepackage{geometry}
\usepackage{float}
\usepackage[backend=biber]{biblatex}
\usepackage{caption}
\usepackage{needspace}
 \usepackage[spanish]{babel}

\lstdefinestyle{json}{
  basicstyle=\ttfamily\footnotesize,
  breaklines=true,
  showstringspaces=false
}

\title{Instalación, configuración y utilización de un nodo Bitcoin en Linux}

\author{
    José Ulloa Araya \\
    \texttt{joseeduardo.ulloa24@estudiantes.uva.es} \\
    ORCID: \href{https://orcid.org/0009-0001-2385-6990}{0009-0001-2385-6990} \\
    \\
    Diego R. Llanos \\
    \texttt{diego.llanos@uva.es} \\
    ORCID: \href{https://orcid.org/0000-0001-6240-9109}{0000-0001-6240-9109} \\
    \\
    Escuela de Ingeniería Informática\\
    Universidad de Valladolid
}

\date{May 2025}

\begin{document}

\maketitle

\begin{abstract}
Este trabajo documenta la instalación, configuración y funcionamiento de un nodo completo de Bitcoin en un entorno Linux, desde la compilación manual del código fuente hasta la sincronización completa con la red. Se exponen las fases técnicas del proceso, se analizan los principales ficheros generados por Bitcoin Core y se estudian de forma empírica los efectos de los parámetros \texttt{txindex}, \texttt{prune}, \texttt{dbcache}, \texttt{maxmempool} y \texttt{maxconnections}. También se documentan los recursos del sistema durante el mecanismo de descarga de bloques (IBD) y se argumenta la importancia operativa de cada uno de ellos.
Este trabajo es una base sólida para futuras propuestas de investigación sobre el rendimiento de los nodos de Bitcoin o para la construcción de herramientas de consulta de datos sobre la blockchain.
\end{abstract}

\bigskip

\textbf{Palabras clave:} Bitcoin Core, blockchain, full node, Linux, initial block download, system performance, configuration parameters.



\section{Introducción}


Bitcoin es una red descentralizada que permite realizar transacciones (sin recurrir a intermediarios) mediante una arquitectura técnica distribuida basada en nodos. Cada nodo Bitcoin completo valida, almacena y propaga bloques y transacciones que se ajustan a las reglas del protocolo, constituyendo en sí mismo una porción de integridad y seguridad del sistema. Conseguir ejecutar un nodo propio no solo permite participar de la red, sino también acceder directamente a los datos de la blockchain, pudiendo así verificar la información sin depender de terceros.

Este trabajo busca facilitar la comprensión del funcionamiento de un nodo Bitcoin completo en un entorno Linux y documentar el proceso de instalación, configuración y funcionamiento del software Bitcoin Core.

En cuanto al objetivo general del trabajo, se ha querido documentar detalladamente el proceso, poniendo el acento en los archivos generados durante la ejecución, así como en los efectos que tendrán los diferentes parámetros de configuración sobre el rendimiento y la operatividad del nodo. Con ello se espera analizar el sistema de archivos que produce Bitcoin Core; evaluar configuraciones relevantes, como txindex, prune, dbcache, maxmempool y maxconnections; y generar una guía sobre la instalación y la ejecución básica del nodo en entornos domésticos. Entre las variables objeto de observación se tienen en cuenta el uso de los recursos del sistema, los tiempos de sincronización, así como las capacidades de consulta y análisis que trae consigo cada configuración.

En síntesis, este trabajo muestra que es posible instalar y operar un nodo completo de Bitcoin en un entorno local empleando recursos de la máquina moderados, siempre y cuando los ajustes correspondientes se configuren correctamente y se tengan claras las consecuencias de las distintas opciones de configuración.

\section{Qué es un nodo Bitcoin y cuál es su función en la red}

Lo primero es entender que Bitcoin es una red descentralizada, que tiene la finalidad de permitir la posibilidad de transferir valor entre pares, sin ningún tipo de intermediario. En donde los nodos son el elemento clave dentro de este sistema, pues son los encargados de la validación de reglas del protocolo, además del almacenamiento de todo el historial de la cadena de bloques, así como de propagar la información al resto de los usuarios. En el libro \textit{Criptomonedas para Dummies}, de Carlos Callejo y Víctor Ronco, nos dice que "cada uno de estos equipos informáticos es un nodo”, y que “todos se encuentran conectados entre sí formando una sólida red en la que cada uno guarda una copia parcial o completa del historial de transacciones”\parencite[p.~18]{ronco2020}. Es decir, en simples palabras, son computadoras conectadas entre sí, que conforman una red en la cual cada una mantiene una copia.

Incluso Andreas M. Antonopoulos, en el libro \textit{Mastering Bitcoin}, nos indica que un nodo completo no sólo almacena la cadena de bloques, sino que “participa activamente validando las transacciones y los bloques, asegurando la integridad y la coherencia de la red”\parencite[p.~141]{antonopoulos2019}. Por lo tanto, cada computadora (nodo) tiene la capacidad de rechazar un bloque o un conjunto de bloques que no están de acuerdo a las normas, incluso si ha sido construido por un actor con gran poder de cómputo.

\begin{figure}[H]
\centering
\includegraphics[width=0.9\textwidth]{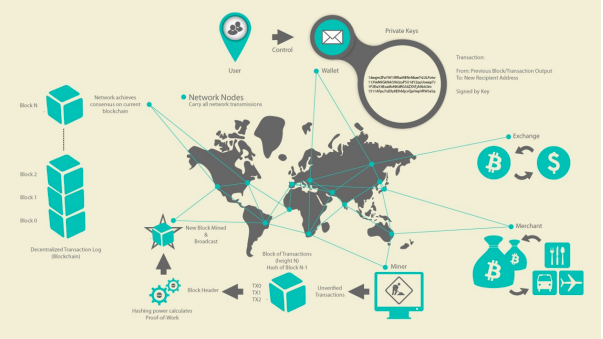}
\vspace{-0.7em}
\caption{Esquema visual del funcionamiento de la red Bitcoin. Infografía publicada en \textit{Mastering Bitcoin}, Antonopoulos (2019), p.~24.}
\label{fig:arquitectura-bitcoin}
\end{figure}

En la \texttt{Figura~\ref{fig:arquitectura-bitcoin}} se puede apreciar un esquema visual del funcionamiento de la red Bitcoin. Se observan usuarios, nodos, mineros, billeteras y el flujo de transacciones. 

\subsection{Un ejemplo simple para entender: boletín de notas de un colegio.}

A lo mejor el lenguaje técnico anterior puede sonar bastante complicado, por lo cual, mediante este ejemplo se busca mejorar la comprensión de lo que es Bitcoin: pensemos en un sistema escolar, en donde todos los cursos tienen un profesor y cada profesor lleva una copia idéntica del boletín de calificaciones de todos los alumnos. Si un profesor quiere anotar algo en el boletín de notas, debe primero validarse con el resto de profesores. Tan pronto como todos los profesores llegan a un acuerdo en que el estudiante efectivamente ha entregado el examen y que la nota que se propone es válida, entonces la nota se anota en el boletín; en otro caso, si por ejemplo existe alguna discrepancia en el boletín y no parece convincente para el resto de los profesores, el cambio en la nota se niega. Así mismo actúan los nodos en la red de Bitcoin, es decir, se validan los unos con los otros para validar un libro contable común\footnote{En el contexto de Bitcoin, cuando decimos "libro común contable", nos referimos al blockchain}, sin la necesidad de una autoridad central.

\subsection{Propiedades fundamentales de los nodos}

En \textit{El pequeño libro de Bitcoin}, se nos indica que “hay miles de validadores en todo el mundo que verifican la cadena de bloques de Bitcoin y almacenan el historial completo de las transacciones de bitcoin. Estos validadores se denominan "nodos completos”\parencite[p.~84]{barrera2019}, por lo tanto, en simples palabras, estos nodos completos garantizan la descentralización (puesto que no hay un único punto de control), la seguridad (debido a que la verificación de los datos se lleva a cabo a través de un proceso distribuido), la transparencia (en donde todas las partes pueden comprobar el historial público de las transacciones) y la resistencia a la censura (ningún nodo puede imponer cambios en caso de no existir consenso).

\subsection{Distinción entre nodos y mineros}

Esta puede ser una pregunta usual, ya que algunas personas llegan a pensar que todos los nodos son también mineros, pero nada más lejos de la realidad, ya que los mineros son los que se encargan de proponer nuevos bloques, así como competir por las recompensas de validación, mientras que, por otro lado, los nodos completos validan que los bloques propuestos cumplan con las reglas del protocolo. En \textit{Criptomonedas para Dummies} se aclara bastante bien esta diferencia al decir que "El nodo solo almacena información; el minero 'trabaja' para validar transacciones.”\parencite[p.~21]{ronco2020}.

\subsection{Por qué ejecutar un nodo completo}

Poner en funcionamiento un nodo completo de Bitcoin es una manera real de participar en la soberanía digital; en este sentido, en \textit{El pequeño libro de Bitcoin} se nos indica que “cada usuario puede, al ejecutar un nodo completo, verificar todo y no tener que confiar en nadie más”\parencite[p.~98-99]{barrera2019}. Pues, por la facilidad técnica actual para instalar y mantener un nodo en casa, hasta los usuarios menos expertos pueden reforzar la red y verificar mecánicamente sus propios saldos y transacciones, sin tener que recurrir a terceros. Cosa que se busca demostrar en este paper.

\subsection{Bitcoin como infraestructura distribuida}

Por último, en el libro \textit{Mastering Bitcoin} se pone de manifiesto que “Bitcoin es también el nombre del protocolo, una red entre pares y una innovación de computación distribuida”\parencite[p.~13]{antonopoulos2019}. Cada nodo es parte de esa red, la existencia de la cual garantiza el hecho de que Bitcoin siga siendo lo que siempre ha pretendido ser: un sistema de dinero electrónico sin control central.
\section{Instalación y configuración de Bitcoin Core en un sistema Linux}

Instalar un nodo completo de Bitcoin en una máquina GNU/Linux forma parte de una experiencia formativa indispensable para entender el funcionamiento de esta red descentralizada. Lejos de ser una mera ejecución de un software, esta instalación es un proceso que lleva consigo dilemas técnicos que impactan directamente sobre la robustez, seguridad y extensibilidad del sistema resultante. Aquí documentamos la instalación de \texttt{Bitcoin Core} versión 29.0, que se compila a partir del código fuente y se configura como servicio del sistema.

\subsection{Razones para la compilación manual}

Si bien, ya existiendo versiones compiladas de Bitcoin Core, se optó por compilarlo a partir de código fuente y no de una versión precompilada por tres razones fundamentales:
\begin{itemize}
    \item Permite saber y controlar exactamente cuáles son las opciones de configuración habilitadas, por ejemplo, el soporte para billetera (wallet), o la base de datos de transacciones completas (txindex)\footnote{Esta opción permite consultar transacciones antiguas usando solo su txid}, o de avisos mediante ZeroMQ\footnote{ZeroMQ es una biblioteca de mensajería asincrónica utilizada en Bitcoin Core para emitir notificaciones en tiempo real sobre eventos; para más detalle se puede consultar la documentación oficial en \url{https://github.com/bitcoin/bitcoin/blob/master/doc/zmq.md} o ocupar \texttt{bitcoind --help} en la terminal.}
    \item Permite una compatibilidad más completa con el sistema operativo, evitando dependencias con binarios precompilados.
    \item Está en línea con el objetivo del proyecto para investigar la arquitectura de la aplicación Bitcoin a partir de lo básico.
\end{itemize}

\subsection{Preparación del sistema y entorno dedicado}

Para ello se ha trabajado en un equipo con procesador Intel Core i5-8400, con 6 núcleos y 16 GB de RAM, un disco de 2 TB a disposición para la cadena de bloques, montando parte del sistema en el directorio \textit{/opt/BLOCKCHAIN}, con permisos exclusivos para el usuario bitcoin. Este aislamiento tiene por objetivo:
\begin{itemize}
    \item Facilitar la gestión del volumen de datos (actualmente más de 500 GB).
    \item Mejorar el rendimiento en lectura/escritura al separar blockchain del sistema operativo.
    \item Garantizar la futura migración y/o respaldo del nodo.
\end{itemize}

\subsection{Descarga y compilación del código fuente}

Se optó por \textit{bitcoin v29.0}, la cual fue descargada directamente (y sin ningún tipo de modificación) desde el repositorio oficial en GitHub. Se utilizaron las herramientas de construcción basadas en CMake, de acuerdo con la última documentación oficial del repositorio de bitcoin:

\begin{verbatim}
    $ git clone https://github.com/bitcoin/bitcoin.git
    $ cd bitcoin
    $ git checkout v29.0
\end{verbatim}

Se instalaron las dependencias necesarias, priorizando los paquetes que están disponibles en los repositorios oficiales de Ubuntu:

\begin{verbatim}
    $ sudo apt install build-essential cmake pkgconf python3 libevent-dev \ 
    libboost-dev libsqlite3-dev libzmq3-dev
\end{verbatim}

La configuración del entorno de construcción se realizó mediante:

\begin{verbatim}
    $ cmake -B build -DWITH_SQLITE=ON -DENABLE_WALLET=ON -DWITH_ZMQ=ON
    $ cmake --build build -j$(nproc)
    $ sudo cmake --install build
\end{verbatim}

Con todas estas opciones se habilitó el soporte para billeteras con base de datos SQLite, soporte con bitcoin-cli y soporte para acceso a los eventos ZMQ.

\subsection{Configuración del nodo y seguridad}

Se creó un fichero de configuración en /etc/bitcoin/bitcoin.conf con los parámetros que se aprecian en la \textit{Figura~\ref{fig:parametros}}.

\begin{figure}[H]
\centering
\begin{tcolorbox}[colback=gray!5, colframe=black!40, boxrule=0.5pt, width=0.9\linewidth, enhanced jigsaw]
\begin{verbatim}
txindex=1
datadir=/opt/BLOCKCHAIN
rpcauth=bitcoin:<hash generado>
\end{verbatim}
\end{tcolorbox}
\vspace{-0.7em}
\caption{Configuración de \texttt{/etc/bitcoin/bitcoin.conf}}
\label{fig:parametros}
\end{figure}

La opción \textit{txindex=1} es importante, pues permite realizar consultas sobre cualquier transacción que se encuentre en la cadena y no sólo sobre aquellas que se encuentren asociadas a direcciones propias\parencite[p.~56]{antonopoulos2019}, lo cual es muy importante en el caso de que se deseen desarrollar posteriormente herramientas de visualización.

El campo rpcauth fue generado con el script \textit{rpcauth.py} (se encuentra incluido dentro del paquete bitcoin descargado de Github), el cual permite autenticación segura por RPC para las herramientas analíticas.

\subsection{Ejecución como servicio del sistema}

A fin de asegurar su persistencia y facilidad de administración, el nodo fue configurado como servicio systemd, por lo cual se creó el fichero con la configuración que se aprecia en la \textit{Figura~\ref{fig:systemd}}.

\begin{figure}[H]
\centering
\begin{tcolorbox}[colback=gray!5, colframe=black!40, boxrule=0.5pt, width=0.9\linewidth, enhanced jigsaw]
\begin{verbatim}
[Unit]
Description=Bitcoin daemon
After=network.target

[Service]
ExecStart=/usr/local/bin/bitcoind -conf=/etc/bitcoin/bitcoin.conf \
  -pid=/run/bitcoind/bitcoind.pid
User=bitcoin
Group=bitcoin
Type=forking
PIDFile=/run/bitcoind/bitcoind.pid
Restart=on-failure
RuntimeDirectory=bitcoind

[Install]
WantedBy=multi-user.target
\end{verbatim}
\end{tcolorbox}
\vspace{-0.7em}
\caption{Configuración de \texttt{Bitcoin daemon}}
\label{fig:systemd}
\end{figure}

Con el fin de iniciar el nodo automáticamente al arrancar el sistema. Además, permitirá reiniciarlo fácilmente ante cualquier error, e incluso administrarlo mediante \texttt{systemctl}.

\subsection{Validación inicial}

El nodo fue ejecutado correctamente y comenzó su sincronización con la red de Bitcoin. Se verificó si estaba funcionando correctamente, el siguiente comando:

\begin{verbatim}
    $ sudo systemctl status bitcoind
\end{verbatim}

Y para verificar el estado de sincronización, este comando:
\begin{verbatim}
    $ bitcoin-cli getblockchaininfo
\end{verbatim}

ELa sincronización completa de la cadena de bloques, se realizó en unas 23 horas, lo que se detalle mas profundamente en el siguiente capítulo, 
\section{Descripción del proceso de sincronización}

Con la configuración ya establecida en el capítulo anterior, y una vez ejecutado el nodo como un servicio de sistema mediante \texttt{systemd}, se inició la descarga de los bloques de la red Bitcoin (proceso conocido como Initial Block Download), el cual se encarga de descargar y verificar uno a uno los nodos, lo cual permite que la máquina bitcoin funcione de manera íntegra y autónoma.

El nodo tuvo una sincronización desde cero y, para controlar el progreso de la manera más exacta posible, se creó un pequeño script en bash que cada 30 minutos consultaba al cliente mediante \texttt{bitcoin-cli} y almacenaba en un CSV la altura del bloque y el valor de \texttt{verificationprogress}, que es el porcentaje de progreso de la validación de la cadena.

El procesamiento comenzó el 29 de abril a las 05:54, desde el bloque 304263. Cuando se evaluaron los resultados el 30 de abril a las 05:00, la altura alcanzada era de 894549, decidiéndose en ese momento parar la ejecución del script, lo que significa descargar y validar 590.000 bloques en aproximadamente 23 horas.

\subsection{Evolución y resultados del proceso}

\begin{figure}[H]
    \centering
    \includegraphics[width=0.95\textwidth]{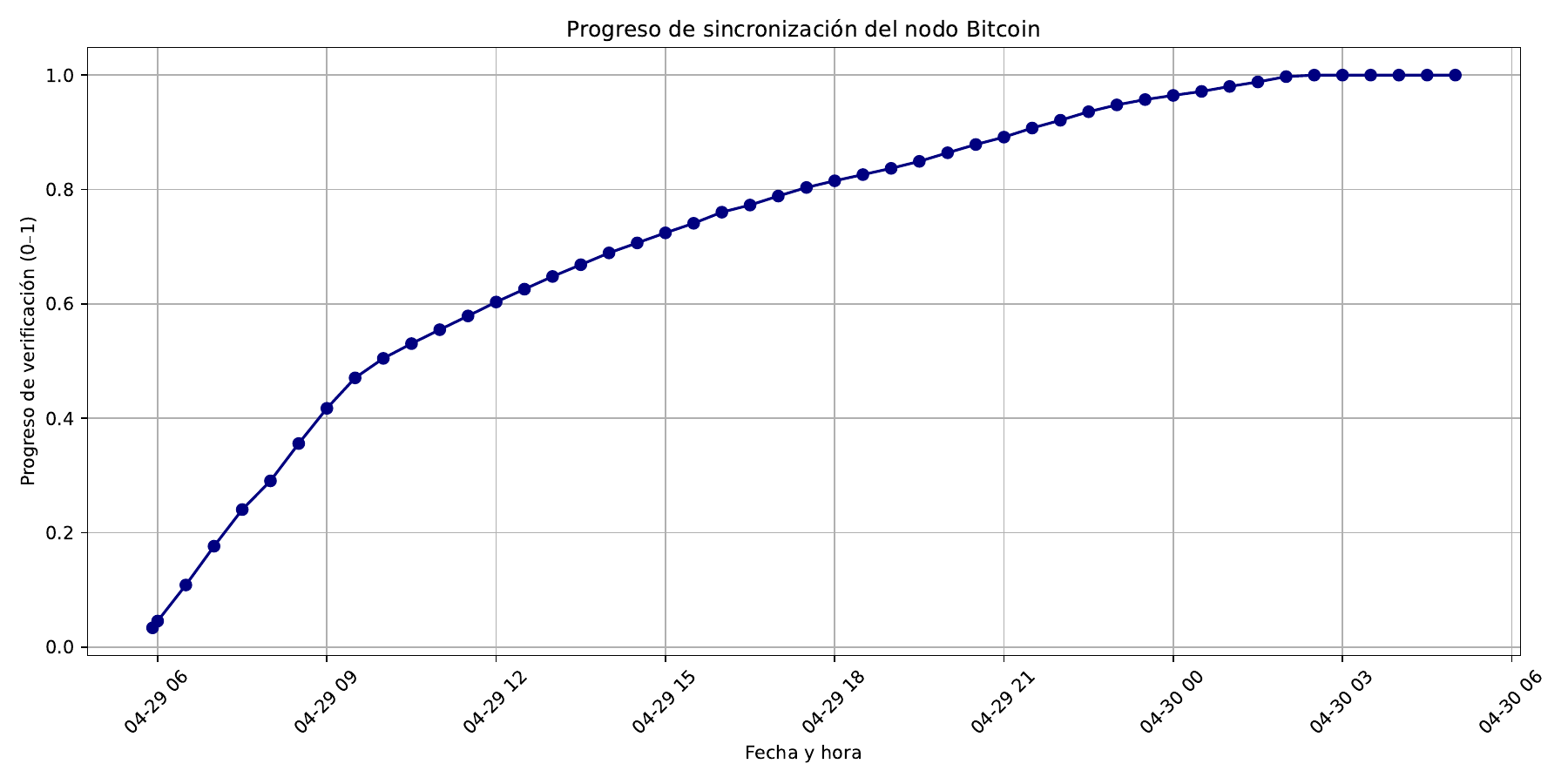}
    \caption{Progreso de sincronización del nodo Bitcoin durante el proceso de descarga inicial.}
    \label{fig:sincronizacion-grafico}
\end{figure}

Durante el proceso, la variable verificationprogress mostró un aumento continuo y sostenido, tal como se puede apreciar en la \texttt{Figura Figura~\ref{fig:sincronizacion-grafico}}. Dicha variable puede tomar valores dentro del intervalo de 0 a 1, indicando el avance de la sincronización mediante una estimación que tiene en cuenta la actividad reciente de la red junto con el historial validado.

Las primeras horas transcurrieron sin interrupción, y tal como se puede apreciar en la \texttt{Figura Figura~\ref{fig:sincronizacion-grafico}}, se ve un continuo incremento de la frecuencia de descarga y de validación de bloques, en donde para la madrugada del 30 de abril, el valor de verificationprogress ya alcanzaba casi 1, aunque la sincronización no se había culminado aún. Recién a las 05:00 se alcanzó la altura total de 894549 bloques y encabezados, condición necesaria para que el nodo culminara el estado de descarga inicial, de acuerdo a la condición inicial de \texttt{inicialblockdownload = false}.

Cosa que se verificó mediante el comando:
\begin{verbatim}
    bitcoin-cli getblockchaininfo | grep initialblockdownload
\end{verbatim}

\subsection{Discusión de eventos relevantes}

Un hecho particular que se da hacia el final de la ejecución es que la verificación de verificationprogress, incluso, presentaba estabilización (o leve oscilación), a pesar de que bloques y encabezados estaban prácticamente equilibrados, lo cual podría generar cierta inquietud, y llegar a pensar que se generó un error en la sincronización, pero esto es acorde al algoritmo de estimación de progreso (dentro del propio Bitcoin Core) que constantemente recalibra el estado de sus estimaciones en función de los bloques más recientes y de la actividad reciente de la red.

El nodo alcanzó su estado final de ejecución satisfaciendo dos condiciones de forma simultánea; por un lado, la igualdad de bloques y encabezados y por el otro, una validación correctiva de la propia cadena de bloques suficientes, con lo cual el sistema consideraba completado el proceso de inicialización, es decir, cuando \texttt{inicialblockdownload = false}.

\subsection{Valor del monitoreo automatizado}

La utilización del script de monitoreo permitió comprobar empíricamente el rendimiento del nodo y mantener un registro con un grado de detalle suficiente del proceso durante un tiempo prudente. Esto no sólo permitió verificar que el sistema estuviera funcionando correctamente, sino que además proporcionó cifras que evidencian que un nodo completo puede sincronizarse en menos de 24 horas bajo unas condiciones adecuadas de hardware, red y de una correcta configuración.

Este tipo de monitoreo resulta especialmente útil en contextos investigativos, porque permite comparar diferentes escenarios, identificar donde pueden producirse cuellos de botella e incluso averiguar el efecto de la implementación de opciones como \texttt{txindex=1} o el modo de \texttt{poda}, aunque, tal como se ha señalado en el tercer capítulo, la sincronización únicamente se realizó con la opción sin poda, para poder almacenar todo el historial de bloques en disco.
\section{Análisis de los archivos generados por Bitcoin Core}

Una vez que ha sido finalizada la sincronización inicial del nodo Bitcoin, procederemos a analizar el directorio de datos indicado en la configuración (en nuestro caso, \textit{datadir=/opt/BLOCKCHAIN}). En dicho directorio se han ido generando una cantidad de archivos y subdirectorios que representan el estado del nodo y los diferentes procesos internos que hacen que el nodo funcione.

Para proceder al examen de los archivos de una forma ordenada, se optó por ejecutar el comando \textit{ls -lhR /opt/BLOCKCHAIN}, que permite la obtención de un listado completo y recursivo de todos los archivos que hayan sido ya generados por Bitcoin Core. Dicho listado fue asumido como el punto de partida que haría posible la identificación, clasificación y estudio de los archivos y carpetas que podrían ser interesantes.

Todas las estructuras, ya sean las de almacenamiento de los bloques y estados de transacciones, hasta registros de debug y parámetros de red, todas tienen alguna función específica. Este capítulo tiene como objetivo examinarlas, fijándose en el objetivo técnico de cada una de ellas.

Los principales archivos y directorios detectados en la ruta \textit{/opt/BLOCKCHAIN} se pueden apreciar en la \textit{Figura~\ref{fig:estructura-blockchain}}.

\begin{figure}[H]
\centering
\begin{tcolorbox}[colback=gray!5, colframe=black!40, boxrule=0.5pt, width=0.9\linewidth]
\begin{verbatim}
/opt/BLOCKCHAIN
|-- banlist.json
|-- blocks/
|   `-- index/
|-- chainstate/
|-- debug.log
|-- fee_estimates.dat
|-- indexes/
|   `-- txindex/
|-- peers.dat
`-- settings.json
\end{verbatim}
\end{tcolorbox}
\vspace{-0.7em}
\caption{Estructura de archivos y subdirectorios en \texttt{/opt/BLOCKCHAIN}.}
\label{fig:estructura-blockchain}
\end{figure}

\subsection{Análisis del archivo \textit{banlist.json}}

El archivo \textit{banlist.json} fue generado y gestionado automáticamente por Bitcoin Core y su finalidad es ir registrando las direcciones IP o rangos de red que han sido bloqueadas por el nodo, ya sea de manera manual o automática (dependiendo de acciones maliciosas o sospechosas). Este mecanismo permite que el nodo esté protegido frente a los ataques o conexiones indeseadas.

Tal como se puede observar en la \textit{Figura~\ref{fig:banlist.json}}, el campo \textit{banned\_nets} se encuentra vacío, lo que indica que el nodo no ha considerado necesario bloquear alguna dirección hasta ahora. Además, la advertencia \textit{\_warning\_} indica que este archivo no debe ser editado manualmente mientras el nodo se ejecuta, pues este cambio podría ser sobreescrito.

\begin{figure}[H]
\centering
\begin{tcolorbox}[colback=gray!5, colframe=black!40, boxrule=0.5pt, width=0.9\linewidth, enhanced jigsaw]
\begin{lstlisting}[style=json]
{
  "_warning_": "This file is automatically generated and updated by Bitcoin Core. Please do not edit this file while the node is running, as any changes might be ignored or overwritten.",
  "banned_nets": []
}
\end{lstlisting}
\end{tcolorbox}
\vspace{-0.7em}
\caption{Contenido \texttt{banlist.json}}
\label{fig:banlist.json}
\end{figure}

Tras ejecutar el comando \textit{stat /opt/BLOCKCHAIN/banlist.json}, se comprobó que el tamaño del archivo es de 221 bytes y fue modificado por última vez el 29 de abril de 2025, con lo cual queda claro que no ha sido modificado desde entonces. Esta información confirma que su estado actual se corresponde con el del archivo base que fue creado por Bitcoin Core en la primera ejecución del nodo.

\subsection{Análisis del directorio \textit{blocks/} y su índice}

Se llevó a cabo el análisis del directorio \textit{/opt/BLOCKCHAIN/blocks}, el cual presenta los archivos binarios de los bloques y su respectivo índice. Para ello, se realizó el listado del contenido de la carpeta principal mediante el comando:
\begin{verbatim}
    $ ls -lh /opt/BLOCKCHAIN/blocks
\end{verbatim}

Luego, se evaluó el espacio total ocupado por esta carpeta mediante el comando:

\begin{verbatim}
    $ du -sh /opt/BLOCKCHAIN/blocks
\end{verbatim}

Que dio como resultado un tamaño de alrededor de \textit{471 GiB}. De modo que se puede confirmar que el nodo habría almacenado íntegramente la cadena de bloques sin utilizar ningún tipo de poda (\textit{prune=0}).

También se analizaron los metadatos de todos los archivos \textit{.dat} mediante el comando:

\begin{verbatim}
    $ stat /opt/BLOCKCHAIN/blocks/*.dat
\end{verbatim}

Esta permitio verificar que los archivos habían sido creados y modificados por el usuario \textit{bitcoin}, con permisos restringidos a dicho usuario (modo \textit{0600}). Las fechas indicaban una creación progresiva, lo cual es consistente con el avance cronológico de la sincronización.

Luego para poder determinar la cantidad exacta de archivos en dicha carpeta, se ejecutó:

\begin{verbatim}
    $ ls /opt/BLOCKCHAIN/blocks | wc -l
\end{verbatim}

Confirmando la existencia de más de 9.000 archivos (entre \texttt{blk} y \texttt{rev}). Finalmente, se procedió a analizar el subdirectorio index/, el cual contiene la base de datos LevelDB (encargada de indexar los bloques para un acceso eficiente):

\begin{verbatim}
    $ ls -lh /opt/BLOCKCHAIN/blocks/index
    $ du -sh /opt/BLOCKCHAIN/blocks/index
    $ file /opt/BLOCKCHAIN/blocks/index/*
\end{verbatim}

Lo cual revelo un conjunto de archivos binarios como lo son \texttt{MANIFEST}, \texttt{LOG} y múltiples ficheros \texttt{*.ldb}, lo que claramente  indicaban que el nodo ha sido construido completo y al mismo tiempo permiten realizar consultas rápidas mediante \texttt{blockhash}.

\subsection{Análisis del directorio \textit{chainstate/} y el conjunto UTXO}

El directorio chainstate, se encarga de almacenar el conjunto de salidas de transacciones no gastadas\footnote{Conjunto UTXO, que representa el saldo disponible en la red.}, tambien conocido como el \texttt{UTXO}, el cual permite validar las nuevas transacciones de forma rápida, asegurando que los inputs provienen de salidas legítimas y no gastadas.\parencite[p.~121]{antonopoulos2019}

La estructura de la carpeta \texttt{chainstate/} esta optimizada mediante el uso de una base de datos LevelDB que mejora, de forma significativa, las operaciones de lectura y escritura del conjunto UTXO. Los archivos que tienen prefijos como \texttt{000003.log}, \texttt{CURRENT}, \texttt{MANIFEST-000002}, entre otros y son gestionados dentro del LevelDB, impidiendo asi cualquier tipo de modificación, pues cualquier alteración puede producir una corrupción de los datos, obligando eventualmente a realizar una reconstrucción del índice UTXO.

Para poder analizar el tamaño actual del conjunto UTXO, se uso:

\begin{verbatim}
    $ bitcoin-cli gettxoutsetinfo
\end{verbatim}

Obteniendo lo que se puede apreciar en la \textit{Figura~\ref{fig:utxo-info}}, lo cual refleja el estado del conjunto UTXO en el bloque 847.254, con un tamaño de la base de datos de unos 447 MB en disco, con más de 82 millones de salidas no gastadas. Ahora, dentro de los valores mostrados, la variable bogosize se usa como un valor teórico que estima el tamaño lógico del conjunto sin compresión; por otro lado, el campo total\_amount representa la cantidad de BTC que representan las salidas no gastadas. En palabras más simples, es como cuando se genera una especie de captura instantánea del estado actual de fondos disponible en la red Bitcoin, en la cual, si se agrega un nuevo bloque, la base de datos será actualizada con nuevas salidas, y en caso de que fuesen gastadas, se eliminan, de modo tal que los nodos puedan operar de forma eficiente, evitando que tengan que ir recorriendo toda la blockchain para verificar las transacciones.

\begin{figure}[H]
\centering
\begin{tcolorbox}[colback=gray!5, colframe=black!40, boxrule=0.5pt, width=0.9\linewidth, enhanced jigsaw]
\begin{lstlisting}[style=json]
{
"height": 847254,
"bestblock": "000000000000000000020de7d6b77d65d3a8370cce7be65f6c123b4c26950140",
"transactions": 89805073,
"txouts": 82454865,
"bogosize": 6169726781,
"hash_serialized_2": "f0e849efca8c9f01d0e3969e0de21b4525a9d85aa8aaee297e1a21e452d4498d",
"disk_size": 447277290,
"total_amount": 19975766.22532526
}
\end{lstlisting}
\end{tcolorbox}
\vspace{-0.7em}
\caption{Estado del conjunto \texttt{UTXO} consultado}
\label{fig:utxo-info}
\end{figure}

\subsection{Análisis del archivo \textit{debug.log}}

El fichero log contiene la totalidad de los eventos que tienen que ver con la inicialización del software, la lectura de la configuración, las conexiones con los peers, la recepción de los bloques y el mantenimiento total del nodo. Por lo cual se considera un recurso clave para realizar chequeos del estado del nodo de forma continua y de este modo, poder detectar posibles fallos.

En el caso del nodo sobre el cual se están haciendo las pruebas actuales, el archivo tiene un tamaño de unos 210 Mb aproximados, con un total de 909.217 líneas, creado el 29 de abril de 2025, y la última modificación fue el día en que se escribió este apartado (23 de mayo de 2025), lo que quiere decir que el nodo ha ido funcionando de forma ininterrumpida desde que fue iniciado.

La \textit{Figura~\ref{fig:debug-head}} muestra un extracto de las primeras líneas del fichero, donde se aprecia claramente una correcta inicialización del nodo (junto a la carga de parámetros desde el archivo de configuración).

\begin{figure}[H]
\centering
\begin{tcolorbox}[colback=gray!5, colframe=black!40, boxrule=0.5pt, width=0.9\linewidth, enhanced jigsaw]
\begin{lstlisting}[basicstyle=\ttfamily\footnotesize]
2025-04-29T03:37:20Z Bitcoin Core version v29.0 (release build)
2025-04-29T03:37:20Z Using the 'sse4(1way),sse41(4way),avx2(8way)' 
SHA256 implementation
2025-04-29T03:37:20Z Using RdSeed as an additional entropy source
2025-04-29T03:37:20Z Using RdRand as an additional entropy source
2025-04-29T03:37:20Z Using data directory /opt/BLOCKCHAIN
2025-04-29T03:37:20Z Config file: /etc/bitcoin/bitcoin.conf
2025-04-29T03:37:20Z Config file arg: daemon="1"
2025-04-29T03:37:20Z Config file arg: txindex="1"
2025-04-29T03:37:20Z Command-line arg: conf="/etc/bitcoin/bitcoin.conf"
2025-04-29T03:37:20Z Using at most 125 automatic connections
\end{lstlisting}
\end{tcolorbox}
\vspace{-0.7em}
\caption{Fragmento inicial del archivo \texttt{debug.log}.}
\label{fig:debug-head}
\end{figure}

En las líneas finales del archivo, tal como se puede apreciar en la \textit{Figura~\ref{fig:debug-tail}}, se confirma que el nodo está completamente sincronizado, recibiendo nuevos bloques y conectándose activamente con otros nodos de la red, además de no presentar errores ni advertencias que indiquen un mal funcionamiento del nodo desde su ejecución.

\begin{figure}[H]
\centering
\begin{tcolorbox}[colback=gray!5, colframe=black!40, boxrule=0.5pt, width=0.9\linewidth, enhanced jigsaw]
\begin{lstlisting}[basicstyle=\ttfamily\footnotesize]
2025-05-23T01:58:06Z UpdateTip: new best=000000000000000000017259...
2025-05-23T02:01:55Z New block-relay-only v1 peer connected
2025-05-23T02:02:33Z Saw new header hash=0000000000000000000076f0...
2025-05-23T02:22:42Z UpdateTip: new best=000000000000000000001cc8...
\end{lstlisting}
\end{tcolorbox}
\vspace{-0.7em}
\caption{Fragmento más reciente del archivo \texttt{debug.log}.}
\label{fig:debug-tail}
\end{figure}

\subsection{Análisis del archivo \textit{fee\_estimates.dat}}

El fichero \textit{fee\_estimates.dat} guarda estadísticas que emplea el nodo para poder estimar las comisiones necesarias para poder confirmar las transacciones en un número determinado de bloques. A pesar de estar almacenado de forma binaria, es decir, que no permite la lectura directa, puede ser gestionado indirectamente mediante \textit{bitcoin-cli}.
Además, para poder tener una visión general de las últimas modificaciones del fichero y su creación, se utilizó el comando:

\begin{verbatim}
    $ stat /opt/BLOCKCHAIN/fee_estimates.dat
\end{verbatim}

Indicando que el archivo tiene un tamaño aproximado de 250 KB. Fue creado el mismo día en que se inició el nodo, y desde entonces no ha sido alterado en su estructura, pero eso no quita que su contenido ha sido actualizado periódicamente por Bitcoin Core a través del tiempo, pero sin que haya habido reinicios, ni regeneraciones del archivo. Esta actividad de actualización fue confirmada al revisar el log \textit{debug.log} mediante el siguiente comando:

\begin{verbatim}
    $ grep fee_estimates.dat /opt/BLOCKCHAIN/debug.log
\end{verbatim}

El resultado obtenido de la información posterior al comando. Nos indica cómo el nodo escribe en este archivo cada una hora aproximadamente, incluyendo entradas que siguen el patrón de \textit{“Flushed fee estimates to fee\_estimates.dat”}, lo que significa que se están almacenando las estimaciones de las entradas en función de la evolución que la red ha tenido recientemente.

Con el objetivo de comprobar la utilidad de este archivo,se empleo el comando bitcoin-cli estimatesmartfee, el cual accede a las estadísticas almacenadas, ofreciendo sugerencias de comisiones, tal como sugieren el número de bloques en el que se pretende confirmar la transacción. Por ejemplo, haciendo:

\begin{verbatim}
    $ bitcoin-cli estimatesmartfee 1
\end{verbatim}

El nodo sugiere una tarifa acorde a una confirmación rápida, dentro de los próximos dos bloques (como si hubiese ejecutado \textit{bitcoin-cli estimatesmartfee 2}), obteniendo el resultado que se puede apreciar en la \textit{Figura~\ref{fig:estimatesmartfee-1}}.

\begin{figure}[H]
\centering
\begin{tcolorbox}[colback=gray!5, colframe=black!40, boxrule=0.5pt, width=0.9\linewidth, enhanced jigsaw]
\begin{lstlisting}
{
  "feerate": 0.00003145,
  "blocks": 2
}
\end{lstlisting}
\end{tcolorbox}
\vspace{-0.7em}
\caption{Comisión primeros 2 bloques.}
\label{fig:estimatesmartfee-1}
\end{figure}

Además, se quiso verificar que el sistema restringe el uso de valores fuera del rango permitido al ejecutar:

\begin{verbatim}
    $ bitcoin-cli estimatesmartfee 0
\end{verbatim}

Lo que probó que el sistema restringe bien los valores fuera de los parámetros permitidos, ya que una llamada de fuera de rango como estimatesmartfee 0 devuelve el mensaje de error indicando que debe estar entre 1 y 1008, tal como se puede ver en la \textit{Figura~\ref{fig:estimatesmartfee-0}}.

\begin{figure}[H]
\centering
\begin{tcolorbox}[colback=gray!5, colframe=black!40, boxrule=0.5pt, width=0.9\linewidth, enhanced jigsaw]
\begin{lstlisting}
error code: -8
error message:
Invalid conf_target, must be between 1 and 1008
\end{lstlisting}
\end{tcolorbox}
\vspace{-0.7em}
\caption{Restricción del uso de valores}
\label{fig:estimatesmartfee-0}
\end{figure}

Por lo tanto, todo esto confirma que \textit{fee\_estimates.dat} es un fichero fundamental para que el nodo mantenga recomendaciones de tarifas actualizadas y acordes a las condiciones actuales de la red.

\subsection{Análisis del directorio \textit{indexes/} y el índice \textit{txindex/}}

La carpeta \textit{indexes/} contiene datos de apoyo que permiten el acceso rápido a datos concretos de la blockchain. En esta carpeta destaca el índice \textit{txindex}, el cual permite recuperar cualquier transacción en función de su txid sin necesidad de saber el bloque que contiene la transacción. Esto se hace al habilitar la opción \textit{txindex=1} a través del archivo bitcoin.conf, de modo que el nodo va a iniciar la creación de una base de datos que almacena de forma adicional todas las transacciones existentes en la cadena; y este índice se encuentra ubicado en el directorio \textit{indexes/txindex/}, permitiendo consultar cualquier transacción utilizando el comando:

\begin{verbatim}
    $ bitcoin-cli getrawtransaction <txid> 1
\end{verbatim}

Esta activación de este índice produce un aumento del consumo de disco, y al mismo tiempo un tiempo de inicio de la primera sincronización mucho mas largo. De la misma manera, su utilización es de suma importancia para herramientas y bloques de exploradores y scripts que requieren acceder de una forma mas eficiente a cualquier transaccion.

La carpeta de índices \textit{indexes/} en el nodo analizado ocupa unos 58 GB, los cuales casi en su totalidad están concentrados en \textit{txindex/}, donde se almacenan además miles de ficheros de tipo .ldb con un tamaño medio de 34 MB cada uno. Esta capacidad de indexación permite además a desarrolladores e investigadores acceder directamente a los datos históricos de la red Bitcoin, lo que permite estudios estadísticos, reconstrucciones de las transacciones y validaciones cruzadas con otras fuentes.

\subsection{Análisis del archivo \textit{peers.dat}}

El fichero \textit{peers.dat} almacena la información persistente acerca de los nodos pares (peers) que el nodo Bitcoin conoce. Es útil porque permite mantener una lista de nodos que el cliente trata de volver a conectar cuando se inicia, y que puede colaborar a conseguir una conexión adecuada al P2P de Bitcoin.

Una vez más, usando el comando \textit{stat}, se analizó el fichero, el cual tiene un tamaño aproximado de 2,6 MB, y también fue creado al iniciar por primera vez el nodo, y desde esa fecha se ha mantenido actualizado sin evidencia de modificaciones externas al analizar \textbf{peers.dat} dentro de \textit{debug.log}.

\begin{verbatim}
    $ stat /opt/BLOCKCHAIN/peers.dat
    $ grep peers.dat /opt/BLOCKCHAIN/debug.log
\end{verbatim}

Para poder analizar el contenido se ocupó el siguiente comando (que devuelve una lista de objetos Json con la información detallada de cada peer conectado):

\begin{verbatim}
    $ bitcoin-cli getpeerinfo | jq .
\end{verbatim}

En la \textit{Figura~\ref{fig:ejemplo-getpeerinfo}} se puede apreciar parte de lo obtenido, en donde se puede identificar, por ejemplo, indicando como resultado que el nodo conserva conexión con 10 peers. Cada entrada contiene información: id (identificador local asignado por el nodo a cada conexión), addr (dirección IP y puerto del peer remoto), network (tipo de red usada por el peer), servicesnames (capacidades del peer), relaytxes (indica si el peer retransmite transacciones), pingtime y minping (útil para evaluar la latencia de red), subver (versión del software que ejecuta el peer), bytessent y bytesrecv (volumen acumulado de datos enviados y recibidos, en bytes) y connection\_type (tipo de conexión establecida).

\begin{figure}[H]
\centering
\begin{tcolorbox}[colback=gray!5, colframe=black!40, boxrule=0.5pt, width=0.9\linewidth, enhanced jigsaw]
\begin{lstlisting}
[
  {
    "id": 126,
    "addr": "66.37.25.18:8333",
    "addrbind": "157.88.123.154:55228",
    "addrlocal": "157.88.123.154:55228",
    "network": "ipv4",
    "services": "0000000000000c09",
    "servicesnames": [
      "NETWORK",
      "WITNESS",
      "NETWORK_LIMITED",
      "P2P_V2"
    ],
    ...,
    "bytessent_per_msg": {
      ...
    },
    "bytesrecv_per_msg": {
      ...
    },
    ...
  },
  ...
\end{lstlisting}
\end{tcolorbox}
\vspace{-0.7em}
\caption{Ejemplo resumido de primer caso obtenido con \textbf{getpeerinfo}}
\label{fig:ejemplo-getpeerinfo}
\end{figure}

Esta organización estructural se repite para cada uno de los diez peers que se encuentran activamente conectados y también nos permite tener una caracterización de forma integral de la red a la que se encuentra conectado el nodo y la posibilidad de detectar anomalías o cuellos de botella.

Por lo tanto, tras esta evaluación mediante getpeerinfo, se comprueba que el nodo está bien integrado a la red P2P de Bitcoin, manteniendo conexiones con pares que utilizan versiones recientes del protocolo y con capacidades de retransmisión activadas. Este comando tambien permite detectar problemas de conectividad, latencias elevadas o peers desactualizados. Por lo tanto es una herramienta clave para el monitoreo y mantenimiento del nodo Bitcoin.

\subsection{Análisis del archivo \textit{settings.json}}

El archivo \textit{settings.json}, funciona como un archivo de configuración dinámica, que se irá actualizando durante el funcionamiento, pero a diferencia de los archivos de configuración estática, como por ejemplo bitcoin.conf, este archivo se utiliza para almacenar los ajustes que se han cambiado desde las interfaces como bitcoin-qt o con las llamadas RPC\footnote{Remote Procedure Call: Permite controlar el nodo Bitcoin desde otros programas o script}  y que deben persistir entre sesiones.

\begin{verbatim}
    $ stat /opt/BLOCKCHAIN/settings.json
\end{verbatim}

Se usó nuevamente el comando stat, con lo cual se detecta que el tamaño del fichero es cercano a \textit{F190 bytes}. El fichero no se ha modificado nunca desde su creación, y su modo de permisos \textit{F0600} limita el acceso únicamente al usuario propietario (bitcoin), lo cual refuerza la seguridad.

Al ejecutar el comando:
\begin{verbatim}
    $ cat /opt/BLOCKCHAIN/settings.json
\end{verbatim}

Se obtuvo como resultado lo que se puede apreciar en la \textit{Figura~\ref{fig:settings.json}}, en donde se puede apreciar que el archivo solo contiene una advertencia y no tiene ajustes para persistir configurables. Aun así, su mera presencia indica que el sistema está disponible para el almacenamiento de parámetros de forma dinámica en un futuro, sin tener que editar de forma manual el archivo de configuración por defecto.

\begin{figure}[H]
\centering
\begin{tcolorbox}[colback=gray!5, colframe=black!40, boxrule=0.5pt, width=0.9\linewidth, enhanced jigsaw]
\begin{lstlisting}[style=json]
{
"warning": "This file is automatically generated and updated by Bitcoin Core. Please do not edit this file while the node is running, as any changes might be ignored or overwritten."
}
\end{lstlisting}
\end{tcolorbox}
\vspace{-0.7em}
\caption{Contenido de \texttt{settings.json} en el directorio de datos}
\label{fig:settings.json}
\end{figure}
\section{Efectos de ajustes en la configuración}

Con la finalidad de comprobar de forma empírica los efectos producidos por ciertos parámetros de configuración sobre el comportamiento del nodo Bitcoin, se ha decidido realizar unas pruebas controladas y progresivas, las cuales han de ser expuestas más adelante, cada una de ellas aplicada sobre el mismo nodo en ejecución con las configuraciones modificadas unas tras otra, y medidas tomadas inmediatamente antes y después de cada modificación.

Debido a incompatibilidades técnicas entre ciertos parámetros (por ejemplo, txindex=1 no es compatible con prune>0), se optó por seguir un orden de pruebas para evaluar cada parámetro en un contexto coherente, sin tener que reinstalar o reindexar completamente el nodo. Por esta razón, el primer cambio que se va a documentar será la desactivación del índice de transacciones (txindex=0) pero que mantiene aún el histórico de bloques en disco. A partir de ahí, se irá activando el modo de poda y luego otros parámetros adicionales, todos ellos sobre el estado anterior.

Con el objetivo de verificar los efectos específicos de las configuraciones consideradas, sobre todo en lo correspondiente al acceso a transacciones antiguas, se seleccionó manualmente un txid de transacciones antiguas, a través de blockstream.info, siendo la transacción seleccionada la siguiente:

\begin{verbatim}
    8c14f0db3df150123e6f3dbbf30f8b955a8249b62ac1d1ff16284aefa3d06d87
\end{verbatim}

Correspondiente al bloque \#100000:

\begin{verbatim}
    000000000003ba27aa200b1cecaad478d2b00432346c3f1f3986da1afd33e506
\end{verbatim}
    
Y al mismo tiempo una transacción actual: 
\begin{verbatim}
    000000000000000000002e856a0e776fa10c2c2dd061947d72d50b1b30d833f3
\end{verbatim}

Correspondiente al bloque \#898576:

\begin{verbatim}
    000000000000000000002e856a0e776fa10c2c2dd061947d72d50b1b30d833f3
\end{verbatim}

Siendo estas transacciones verdaderamente útiles, una se encuentra lo suficientemente distante en la cadena como para eliminarla de la poda y la otra lo suficientemente cerca como para no verse afectada, lo que permite comparar su disponibilidad en diferentes configuraciones.

\subsection{Desactivación del índice de transacciones: \texttt{txindex=0}}

La primera prueba que se realizó fue la desactivación del índice de las transacciones, modificando el archivo de configuración del nodo, de \texttt{txindex=1} a \texttt{txindex=0}, ya que con esto, el nodo ya no llevará un índice global de todas las transacciones de la blockchain, afectando de esta forma la posibilidad de consultar la información de una transacción por medio de su txid.

Una vez que el archivo \texttt{bitcoin.conf} había sido modificado y el nodo había sido reiniciado, se comprobó que el índice estaba desactivado ejecutando: 

\begin{verbatim}
    $ bitcoin-cli getindexinfo
\end{verbatim}

El resultado que obtuvo fue un objeto vacío, confirmando que no había sido activado ningún índice. Luego, se intentó consultar una transacción antigua, perteneciente al bloque \texttt{\#100000}:
\begin{verbatim}
    8c14f0db3df150123e6f3dbbf30f8b955a8249b62ac1d1ff16284aefa3d06d87
\end{verbatim}
Y una transacción reciente, perteneciente al bloque \texttt{\#898576}:
\begin{verbatim}
    cf0d196a663995487447e67cc1794314c3fba433ba751cc0fee5dcc6c96a0910
\end{verbatim}

Mediante el comando:

\begin{verbatim}
    $ bitcoin-cli getrawtransaction <txid> 1
\end{verbatim}

Obteniendo el siguiente mensaje de error, en ambos casos:

\begin{lstlisting}
error code: -5
error message:
No such mempool transaction. Use -txindex or provide a block hash...
\end{lstlisting}

Este comportamiento confirma que el nodo no puede encontrar una transacción únicamente con su \texttt{txid}. Pero al repetir la consulta y añadiendo el \texttt{blockhash} correspondiente a cada transacción, el nodo ya ha respondido correctamente:

\begin{lstlisting}[language=bash]
bitcoin-cli getrawtransaction 8c14f0... 1 "000000000003ba27aa200b1c..."
bitcoin-cli getrawtransaction cf0d19... 1 "000000000000000000002e8..."
\end{lstlisting}

De lo que se puede concluir que, a pesar de que el índice global esté inactivo, se puede acceder a las transacciones de bloques guardadas en disco (siempre y cuando se tenga constancia del id específico del bloque). Por lo tanto, en esté sentido, la desactivación de txindex reduce el uso de disco y permite que la inicialización del nodo sea mucho más rápida, aunque claramente limita la posibilidad de una exploración más simple de las transacciones en la cadena de bloques.

\subsection{Activación del modo de poda: \texttt{prune=550}}

El método de poda (\texttt{pruning})\footnote{Pruning elimina bloques antiguos del disco para ahorrar espacio, pero no invalida que el nodo siga siendo válido.} permite reducir el espacio ocupado en el disco. Para ello se modificó el archivo de configuración bitcoin.conf añadiendo la línea siguiente:

\begin{verbatim}
    prune=550
\end{verbatim}

Este parámetro indica al nodo que solo debe conservar los bloques más recientes y eliminar de forma progresiva los bloques anteriores, manteniendo un almacenamiento máximo de \texttt{550 MiB} en los archivos de bloques. El resto del estado de la cadena (ya sea encabezados, chainstate o UTXO) se mantiene íntegro, lo que permite validar nuevas transacciones sin tener el historial completo en disco.

Una vez guardados los cambios, se detuvo el nodo y se volvió a iniciar el nodo, y al ejecutar inmediatamente (una vez reiniciado el nodo) el comando \texttt{getblockchaininfo}, nos devolvió un error:

\begin{lstlisting}
error code: -28
error message:
Pruning blockstore...
\end{lstlisting}

Esto quería decir que ya estaba en proceso de poda inicial. Esta operación tardó unos minutos, durante los cuales el nodo no estuvo disponible para otras operaciones. Una vez terminada, se contrastó el resultado con:

\begin{verbatim}
    $ bitcoin-cli getblockchaininfo
\end{verbatim}

Que ha devuelto, como se aprecia en la \textit{Figura~\ref{fig:getblockchaininfo-poda}}, lo cual confirma que el nodo efectivamente está en estado prune, ya que ha eliminado los bloques anteriores a la altura indicada en \textit{pruneheight}. 

\begin{figure}[H]
\centering
\begin{tcolorbox}[colback=gray!5, colframe=black!40, boxrule=0.5pt, width=0.9\linewidth, enhanced jigsaw]
\begin{lstlisting}[style=json]
{
  "chain": "main",
  "blocks": 898625,
  "headers": 898625,
  "bestblockhash": "00000000000000000001ff3e2ddb6e47b00d062e166351f3940992b829ec9881",
  "bits": "17025049",
  "target": "0000000000000000000250490000000000000000000000000000000000000000",
  "difficulty": 121658450774825,
  "time": 1748362081,
  "mediantime": 1748359457,
  "verificationprogress": 0.9999985627622666,
  "initialblockdownload": false,
  "chainwork": "0000000000000000000000000000000000000000c6593f90ef747549b89aaa16",
  "size_on_disk": 553096761,
  "pruned": true,
  "pruneheight": 898283,
  "automatic_pruning": true,
  "prune_target_size": 576716800,
  "warnings": [
  ]
}
\end{lstlisting}
\end{tcolorbox}
\vspace{-0.7em}
\caption{Resultado \texttt{getblockchaininfo} en modo \texttt{Poda}}
\label{fig:getblockchaininfo-poda}
\end{figure}

De la misma forma, también se observó una notable reducción del espacio en disco que ocupaba esto:

\begin{verbatim}
    $ du -sh /opt/BLOCKCHAIN/blocks
    # Resultado: 796M
\end{verbatim}

Y el directorio \textit{chainstate/} también ocupaba cerca de 4.2 GB, lo cual se esperaría en un nodo sincronizado sin historia, pero con un estado correcto.

Para ver el efecto real de la poda, se intentó acceder al bloque \textit{\#100000} y a su transacción Coinbase mediante sus identificadores respectivamente, pero ambos comandos devolvieron esté mensaje de error:

\begin{verbatim}
error: Block not available (pruned data)
\end{verbatim}

En cambio, cuando se realizó la misma prueba en el bloque más reciente y en su transacción asociada, estas seguían siendo accesibles, ya que se encontraban dentro del rango no podado. Esto confirma que el nodo conserva los restos de bloques para su uso en el modo de nodo completo de solo registros para la validación actual, pero no para la parte histórica.

\subsection{Tamaño de la caché de base de datos: \texttt{dbcache}}

El parámetro dbcache\footnote{Controla la cantidad de RAM que utiliza el nodo. A mayor valor, mejor es el rendimiento si se tiene hardware potente}, permite establecer la cantidad de memoria RAM reservada para realizar las operaciones internas necesarias para leer y escribir sobre LevelDB; parámetro que influye en la eficiencia del nodo en el momento del arranque, la verificación de recompensas por la verificación de bloques y las consultas sobre el conjunto UTXO.

Para poder evaluar su efecto, se hicieron dos pruebas, una con el valor de defecto (aproximadamente 450--500 MiB) y la segunda fijando esté en 64 MiB, por lo cual se editó el fichero bitcoin.conf, agregando esta nueva opción:

\begin{verbatim}
dbcache=64
\end{verbatim}

El proceso bitcoind inicial utilizaba aproximadamente 7.6 GiB de memoria residente (RES), tal y como aparece en la salida del comando top:

\begin{verbatim}
    RES:   7,6g
    %MEM:  48,9
\end{verbatim}

Además, se ejecutó el comando \texttt{time bitcoin-cli gettxoutsetinfo} (consulta y calcula el estado actual del conjunto UTXO), el cual demoró 1 minuto y 58 segundos:

\begin{verbatim}
    real    1m58,665s
\end{verbatim}

Tras modificar el valor en \texttt{bitcoin.conf} y reiniciar el nodo, se realizó un cambio significativo en el uso de memoria: el proceso bitcoind pasó a ocupar 4.4 GiB de memoria residente:

\begin{verbatim}
    RES:   4,4g
    %MEM:  28,4
\end{verbatim}

Por supuesto que tras reiniciar el servicio el mensaje mostrado fue:

\begin{verbatim}
error code: -28
error message:
Verifying blocks…
\end{verbatim}

Lo cual es esperable, pues se están realizando los cambios correspondientes, así como las pruebas vistas en los capítulos anteriores. Sobre todo, considerando que ahora cuenta con una caché más pequeña (lo que obliga al nodo a realizar una lectura más intensiva desde disco durante la verificación de bloques), pero una vez terminado el proceso, se pudieron realizar las comparaciones correspondientes.

En esté caso, \texttt{gettxoutsetinfo} presenta una diferencia muy ligera, con un tiempo total de 2 minutos y 1 segundo:

\begin{verbatim}
    real    2m1,423s
\end{verbatim}

Pero a pesar de eso, la reducción de la memoria consumida sí es notable, cosa que se puede apreciar en la \texttt{Tabla ~\ref{tab:dbcache_comparativa}} y que claramente es importante cuando se opera en entornos con recursos limitados, o se tienen limitantes de RAM frente a otros procesos que deben correr también en la misma máquina. Aun así, en nodos con más memoria, lo recomendable es usar un valor alto para optimizar los tiempos de respuesta a corto plazo.

\begin{table}[H]
\centering
\begin{tabular}{|l|c|c|}
\hline
\textbf{Parámetro} & \textbf{Antes} (\texttt{dbcache} por defecto) & \textbf{Después} (\texttt{dbcache=64}) \\
\hline
RAM usada (\texttt{RES})        & 7.6 GiB        & 4.4 GiB \\
Tiempo \texttt{gettxoutsetinfo} & 1m58,665s      & 2m1,423s \\
\hline
\end{tabular}
\caption{Comparación del comportamiento del nodo con diferentes valores de \texttt{dbcache}}
\label{tab:dbcache_comparativa}
\end{table}

\subsection{Tamaño máximo del mempool: \texttt{maxmempool}}

El parámetro maxmempool permite definir la memoria RAM (en MiB) asignada para almacenar las transacciones no confirmadas (mempool); es decir, aquellas que no han ingresado en un bloque, pero están a la espera de ser procesadas por los mineros. Para evaluar los efectos de esté parámetro, primero se observó el comportamiento del nodo con el valor por defecto de maxmempool.

Para ello se obtuvo los valores del mempool, como la cantidad de transacciones almacenadas, uso de memoria actual, y el límite máximo permitido al ejecutar el comando:
\begin{verbatim}
    $ bitcoin-cli getmempoolinfo
\end{verbatim}

Además se quería saber la cantidad cuenta el número de transacciones activas en el mempool, así que se ocupó el comando:
\begin{verbatim}
    $ bitcoin-cli getrawmempool | wc -l
\end{verbatim}

Después, se analizó la distribución de memoria del sistema, incluyendo la usada, libre y en caché con el comando:

\begin{verbatim}
    $ free -h: 
\end{verbatim}

Finalmente se monitoreó el consumo de memoria del proceso bitcoind en tiempo real:

\begin{verbatim}
    $ top -p \$(pidof bitcoind)
\end{verbatim}

Algunos de los valores obtenidos importantes para analizar fueron:
\begin{verbatim}
    "usage": 48377424, "maxmempool": 300000000
    Transacciones en mempool: 15.890
    bitcoind: RES 11,5G, \%MEM 73,9
\end{verbatim}

Posteriormente fue actualizado el fichero de configuración del nodo, modificando el valor del parámetro maxmempool a 50 MiB, con la intención de comprobar cómo esta restricción afecta a la gestión de transacciones pendientes:

\begin{verbatim}
    maxmempool=50
\end{verbatim}

 Tras guardar los cambios, el servicio fue reiniciado con:

\begin{verbatim}
    $ sudo systemctl restart bitcoind
\end{verbatim}

A continuación, se realizaron los mismos pasos anteriores para obtener datos tras los cambios, y se obtuvo que las transacciones fueron limitadas a 10.252, dado que el nodo únicamente conservó las transacciones con las comisiones más altas. Por otro lado, el uso de memoria del propio proceso bitcoind pasó de 11.5 GiB a 1.4 GiB, mostrando el impacto que tuvo dicha disminución de maxmempool, tal como se puede apreciar en los valores relevantes que se presentan a continuación:

\begin{verbatim}
    "usage": 44393264, "maxmempool": 50000000
    Transacciones en mempool: 10.252
    bitcoind: RES 1,4G, \%MEM 9,3
\end{verbatim}

Estos resultados permiten concluir que el valor del parámetro maxmempool tiene una influencia importante en el comportamiento del nodo, que consiste en que la restricción de su valor, por una parte disminuye la capacidad de acumular transacciones por parte del nodo y por otra parte, disminuye el uso de RAM, lo que puede resultar conveniente en la operación de nodos en hardware con recursos limitados o con el propósito de observar en pasiva.

\subsection{Número máximo de conexiones: \texttt{maxconnections}}

Para estudiar el comportamiento del nodo respecto de la configuración de la variable maxconnections, que permite establecer el valor máximo de conexiones de red simultáneas que el nodo aceptará, se realizaron algunas pruebas. La primera acción realizada fue ejecutar el siguiente comando para conocer cuántas conexiones activas había en ese momento: 

\begin{verbatim}
    $ bitcoin-cli getnetworkinfo | grep connections
\end{verbatim}

También era importante obtener las líneas de estado de red mucho más detalladas, con el total del número de conexiones (connections), las conexiones entrantes (connections\_in) y las conexiones salientes (connections\_out), para lo cual se ejecutó esté comando:

\begin{verbatim}
    $ bitcoin-cli getpeerinfo | jq '. | length'
\end{verbatim}

Obteniendo estos resultados:

\begin{verbatim}
    "connections": 10,
    "connections_in": 0,
    "connections_out": 10
\end{verbatim}

Los cuales confirman un total de 10 conexiones salientes activas, pero como lo importante es poder comparar el uso del CPU y memoria se ejecutó esté comando:

\begin{verbatim}
    $ top -p $(pidof bitcoind)
\end{verbatim}

Obteniendo estos valores:
\begin{verbatim}
    PID:       403251  
    %CPU:      1,0  
    %MEM:      7,5  
    RES:       1,2 g (memoria residente)
\end{verbatim}

Luego se intentó forzar las conexiones a 20, por lo cual se agregó esté parámetro (al fichero bitcoin.conf):

\begin{verbatim}
    maxconnections=20
\end{verbatim}

Para luego proceder a reiniciar el servicio:

\begin{verbatim}
    $ sudo systemctl restart bitcoind
\end{verbatim}

Sin embargo, el nodo seguía manteniendo 10 conexiones, por lo tanto, quedó claro que el número actual de conexiones no se encuentra limitado por el valor de maxconnections, pero como se intenta comprobar algún efecto, se optó por una reducción explícita del número de conexiones, haciendo el correspondiente cambio de maxconnections de 20 a 5, y posteriormente se reinició, comprobando los valores obtenidos cambiaron a:

\begin{verbatim}
    "connections": 5,
    "connections_in": 0,
    "connections_out": 5
\end{verbatim}

De nuevo se ejecutó el top para verificar los efectos en el uso de los recursos, obteniendo estos otros valores:

\begin{verbatim}
    PID:       403367  
    %CPU:      38,7  
    %MEM:      28,9  
    RES:       4,5g
\end{verbatim}

Aun cuando uno podría llegar a pensar que la disminución de conexiones generaría una reducción en el uso de recursos, en el caso concreto se dio la situación opuesta: tras reducir maxconnections de 10 a 5, el proceso bitcoind pasó a utilizar una cantidad muy elevada de CPU y RAM. Esto puede deberse a que ha entrado en una fase muy intensa, intentando mantener el nodo, o bien buscando peers alternativos activos, o también sincronizando alguna de las estructuras de datos internas tras el restart.

\section{Conclusiones}

Se ha conseguido establecer y operar un nodo completo de Bitcoin en un entorno Linux, desde la compilación del código fuente hasta la sincronización del nodo con la red. Se han documentado minuciosamente las etapas de instalación, de configuración y de uso de un nodo, como también el análisis de los archivos generados por Bitcoin Core; se ha prestado un especial interés en su estructura, su objetivo y su relevancia operativa.

Además, se han evaluado empíricamente para la configuración de distintos parámetros (txindex, prune, blockfilterindex, dbcache, maxmempool, maxconnections) y analizando sus efectos en el comportamiento del nodo, así como en el consumo de recursos y en la funcionalidad disponible, lo cual ha llevado a comprobar que es posible mantener un nodo activo en entornos más caseros, y que, por otro lado, queda suficientemente claro que tal nodo puede mantenerse activo siempre que sus parámetros sean ajustados adecuadamente en función del objetivo del usuario.

Este documento puede ser parte fundamental para trabajos futuros relacionados con la investigación de la blockchain y el desarrollo de herramientas de consulta y análisis sobre datos blockchain. Como trabajo fuguro, se propone la continuidad con el proceso de instalación y configuración del nodo, haciendo la automatización en función de diferentes perfiles de usuario (validadores, analistas, desarrolladores, operadores de Lightning, educadores), lo cual manifiesta soluciones adecuadas a sus necesidades y, por otra parte, además abre la posibilidad de una inspección más detallada en qué diversas configuraciones inciden directamente en el rendimiento del nodo.

\printbibliography

\end{document}